# Characterisation of Irradiation Damage in Fe–3Cr and Fe–5Cr: A Study on the Effects of Chromium Content and Temperature

Chandra Bhusan Yadav[a,*], Andrew J. London[b], Tonči Tadić[c], Ruqing Xu[d], Wenjun Liu[d], Stjepko Fazinic[c], Suchandrima Das[e]

[a] Department of Mechanical Engineering, University of Bristol, Queens Road, Bristol, BS8 1QU, UK

[b] UK Atomic Energy Authority, Culham Campus, Oxfordshire OX14 3DB, UK

[c] Ruđer Bošković Institute, Bijenička c. 54, Zagreb, 10000, Croatia

[d] Advanced Photon Source, Argonne National Lab, 9700 South Cass Avenue, Argonne, IL 60439, USA

[e] Department of Materials Engineering, Indian Institute of Science, CV Raman Road, Bengaluru, Karnataka 560012

Corresponding author email: chandrabhusan.yadav@bristol.ac.uk

## Abstract

Fe–Cr binary alloys serve as simplified model systems to study irradiation damage relevant to fusion structural materials. Here, Fe-3%Cr and Fe-5%Cr samples were irradiated with 4 MeV Fe ions under a dose rate of $4 \times 10^{-5}$ dpa/s across a linear thermal gradient (120°C–480°C) in a single experiment, enabling direct comparison of temperature and Cr content effects under identical conditions. Depth-resolved Laue micro-diffraction (~$10^{-4}$ strain sensitivity), nanoindentation, and AFM reveal non-monotonic evolution of lattice strain and hardness: both decrease with temperature up to ~300°C, then increase beyond. This turning point reflects a shift from enhanced defect mobility and partial recovery to solute-defect clustering and cavity formation, which stabilize damage. Fe-3%Cr shows consistently higher strain and hardening than Fe-5%Cr, especially at lower temperatures. Minimal change in post-indentation pile-up indicates limited softening or localization. These results highlight how Cr content and temperature jointly affect irradiation response, offering new insights into defect evolution in fusion-relevant alloys.

## 1. Introduction

The global push toward sustainable, low-carbon energy sources have made nuclear fusion a central focus of energy research. Fusion offers an abundant and clean energy source with minimal long-lived radioactive waste, producing mainly helium and short-lived isotopes like tritium. Unlike fission which generates hazardous transuranic byproducts [1]. However, realising practical fusion energy remains challenging due to extreme operating conditions and complex material interactions [2–4]. One of the foremost challenges is developing structural materials that withstand the harsh environment inside fusion reactors, including elevated temperatures (~800 °C at plasma-facing components and 500-600 °C at blankets) [2,5], intense thermal loads up to 10 MW/m$^2$ [6–8], plasm-induced erosion and damage from 14 MeV fusion neutrons [9,10]. Radiation damage is especially critical: high-energy neutrons and helium ions displace atoms, producing point defects (vacancies, interstitials) [11–13], dislocation loops, voids [14–16], and helium-vacancy bubbles [17]. Neutron-induced transmutation reactions, especially (n, α) and (n, p) processes, generate helium and hydrogen within iron-based alloys, exacerbating swelling and embrittlement [18]. Understanding how such defects form, evolve, and affect mechanical properties is essential for designing resilient reactor materials. Given these challenges, the search for structural materials that can resist radiation damage while maintaining mechanical integrity is central to fusion reactor development. To investigate radiation damage, two main emulation methods are used, ion implantation and fission-neutron irradiation. Ion implantation is favoured for rapid damage induction without activating the materials, enabling accelerated studies [19,20].

Given these challenges, the search for structural materials that can resist radiation damage while maintaining mechanical integrity is central to fusion reactor development. Among the leading candidates are Reduced Activation Ferritic-Martensitic (RAFM) steels, which have been specifically designed for such extreme environments. The development of Reduced Activation Ferritic Martensitic (RAFM) steels marked a significant advancement in fusion materials research [21,22]. Introduced in the 1980s, RAFM steels are designed to minimise radioactive waste production upon neutron irradiation, enabling shallow burial disposal of decommissioned components [1,23]. These materials are characterised by their resistance to irradiation-induced swelling, high strength, low thermal expansion, and excellent thermal conductivity, making them ideal candidates for plasma-facing and structural components, including tritium-breeding blankets [24–30].

RAFM steels' performance is influenced by several factors, including temperature, irradiation dose, and chromium content. Chromium plays a pivotal role in determining irradiation resistance, affecting defect clustering [31], swelling behaviour [32,33], and irradiation-induced hardening [34,35]. However, excessive chromium content can lead to solute-rich clusters that intensify hardening and reduce ductility [36–38]. Striking a balance in chromium content is essential for optimising the material's performance. The complex microstructure of RAFM steels, which includes various alloying elements, further complicates the study of defect dynamics [39]. To address this, researchers often use simplified binary Fe-Cr alloys as model systems to isolate the fundamental effects of chromium on radiation damage and mechanical properties.

Research on irradiation effects in Fe-Cr alloys has extensively explored chromium content, temperature, and dose-related changes. Studies have demonstrated that chromium can suppress the formation of vacancy type defects, particularly at higher doses, by altering the stability of radiation-induced point defects [31]. Additionally, in the presence of minor alloying elements such as Ni, Si, and P, vacancy-solute clusters may form and significantly contribute to irradiation-induced hardening [36]. While moderate Cr additions improve radiation tolerance, excessive chromium content can lead to irradiation-induced embrittlement, reducing material ductility and fracture resistance [37]. Similarly, irradiation dose plays a vital role in defect formation and evolution. High dose accelerate the accumulation of vacancy clusters and voids, leading to material swelling and hardening [32,34]. Temperature, on the other hand, directly impacts defect mobility and recovery processes that reduce swelling [33]. However, such recovery mechanisms are often accompanied by creep deformation, complicating material performance under extreme conditions [16], [17].

Researchers have developed various advanced characterisation techniques to study irradiation damage in FeCr alloys, given their importance in nuclear reactor materials due to their high resistance to radiation-induced degradation. Transmission Electron Microscopy (TEM) has been instrumental in identifying dislocation loops and vacancy-type defects, revealing a mix of $<100>$ and $\frac{1}{2}<111>$ loops in Fe–Cr alloys and their dependence on chromium content, which impacts radiation resistance [40]. Atom Probe Tomography (APT) has provided atomic-scale insights into irradiation-induced segregation of Cr, Si, and P at grain boundaries and dislocations, interpreting changes that influence corrosion resistance and mechanical properties [41]. Positron Annihilation Spectroscopy (PAS), complemented by Doppler Broadening Spectroscopy (DBS), has detected vacancy clusters and helium bubbles, confirming the formation of helium-vacancy and their retention profiles, particularly in helium-irradiated Fe12%Cr alloys [42]. Additionally, Small-Angle Neutron Scattering (SANS) has characterised Cr-rich precipitates, highlighting their contributions to radiation hardening [43], while Mössbauer Spectroscopy has further confirmed the role of Cr in reducing vacancy mobility compared to other solutes [44]. Together, these techniques provide the essential information of defect behaviour and microstructural evolution under irradiation, aiding in the development of radiation-resistant materials.

Despite these insights, gaps remain in the understanding of how chromium content, temperature, and irradiation dose collectively influence the behaviour of Fe-Cr alloys. Most studies focus on individual parameters, making it difficult to establish systematic correlations between these factors. Addressing this gap requires a comprehensive investigation that considers the combined effects of chromium content, dose rate, and temperature on defect dynamics and mechanical properties. In this study, we address this by characterising Fe-Cr binary alloy samples with varying chromium content that were irradiated using a novel thermal gradient method [45]. A comprehensive characterisation is employed using a combination of complementary techniques such as synchrotron Laue diffraction, nanoindentation, and atomic force microscopy (AFM). Synchrotron Laue diffraction is chosen for its ability to measure small strains non-invasively with high precision, as demonstrated in studies on helium-irradiated tungsten and under other irradiated materials [46]. Lattice strain measurements of the order of $10^{-4}$ provide detailed insights into the microstructural changes induced by irradiation. Nanoindentation serves as a unique tool to induce and measure deformation in the micron-thin irradiated layer, while atomic force microscopy (AFM) can be used to capture, in high-resolution, the resulting surface variations.

## 2. Materials and methodology
### 2.1. Sample preparation

High-purity polycrystalline Fe-Cr alloys with 3 and 5 wt% chromium content were procured by the European Fusion Development Agreement (EFDA) under the contract named EFDA-06-1901 [47]. These alloys were prepared by mixing of a melted Fe-Cr mother alloy and purified Fe, to achieve a desired chromium concentration. Hot forming was performed at 1150°C to obtain a 20 mm diameter of hexagonal bar, followed by a cold forming with a reduction ratio of 70% to achieve a size of 11 mm diameter. The cold formed bars were heat treated for an hour between 650°C to 850°C under argon to prevent oxidation followed by air cooling. Details of the materials are available in **Table 1**.

**Table 1.**
Chemical composition of Fe3Cr and Fe5Cr alloys.

| Alloy | Cr (wt%) | C (ppm) | S (ppm) | N (ppm) | O (ppm) | P (ppm) |
|-------|----------|---------|---------|---------|---------|---------|
| Fe3Cr | 3.05 | 4 | 3 | 2 | 4 | 5 |
| Fe5Cr | 5.40 | 4 | 3 | 2 | 4 | 5 |

Hereafter, the two alloys containing 3 and 5 wt% Cr will be referred to as Fe3Cr and Fe5Cr, respectively. Bars were cut using an electro-discharge machine in a matchstick shape having size of $1\times1\times25$ mm$^3$.

Surface of the samples were ground using grades P360 through P1200 of SiC abrasive paper for 3-4 minutes depending on surface finish achieved. Subsequently samples were polished by 9, 3, and 1 µm diamond suspension followed by 0.05 µm colloidal silica to obtain a mirror finish. An optical microscope was employed for visual inspection after each grinding and polishing operations to ensure there are no surface deformation present due to previous steps.

## 2.2. Ion implantation

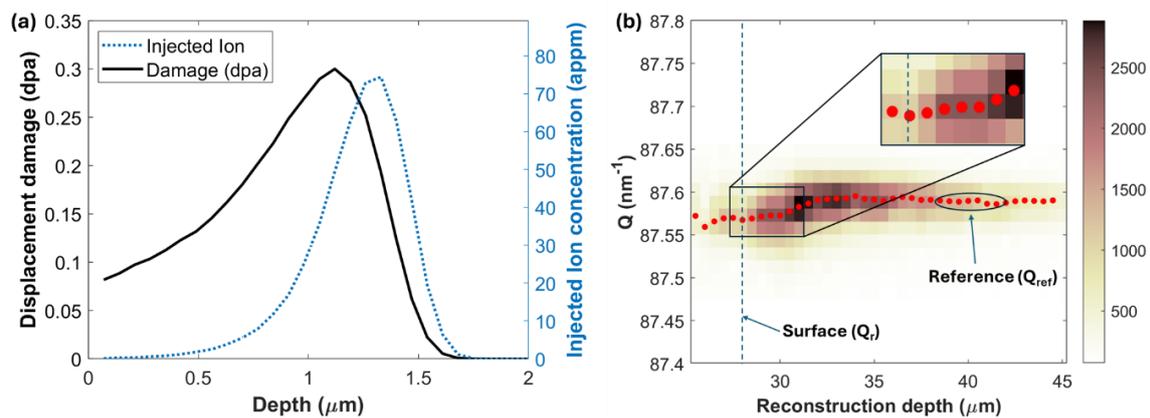

**Fig. 1.** (a) Profile of injected ion represented as dotted curve and damage (solid curve) as a function of depth in the sample, (b) Integrated diffracted intensity (as a function of scattering vector magnitude |Q| and depth in the sample) from a <001> oriented grain in the Fe5Cr sample at 440°C.

Ion implantation was accomplished by Dual Beam Ion Irradiation Facility for Fusion Materials (DiFU) at Ruđer Bošković Institute, Croatia. Samples were irradiated using the TGAS equipment, which is detailed elsewhere [48]. In summary, samples are clamped with a heater at one end and a water-cooled clamp at the other. Two K-type thermocouples were inserted to shallow holes (10 mm separation), 7.5 mm from each end, to measure the sample temperature variation. A temperature gradient of 120 to 480°C was achieved between the two holes on samples. Samples were irradiated with 4 MeV Fe ions with $3\times10^{14}$ ions/cm$^2$ for 6840 seconds, to give a peak damage rate of $4\times10^{-5}$ dpa/s. **Fig. 1** (a) illustrates the injected ions and displacements per atom as a function of depth in the sample. The ion dose required to reach the peak damage level of 0.3 dpa was estimated using SRIM 2013 (quick Kinchin-pease model calculation) using a threshold displacement energy of 40 eV [49]. An oval shaped (approximately 10 mm tall and 4 mm wide) heavy ion beam was rastered over samples, with a frequency of 488Hz, to achieve a uniform irradiated area of $14\times14.5$ mm$^2$. The scanned area was defined by vertical and horizontal slits.

After irradiation, both side of the exposed area of samples were scored with a knife to demarcate the non-irradiated and irradiated regions.

## 2.3. Synchrotron Laue X-ray diffraction

Irradiation damage can introduce defects in the crystal, and these defects create a strain field which can be measured. Micro-beam Laue diffraction can be used to non-destructively probe any lattice strains present in the sample after irradiation with high spatial resolution. The beamline setup 34-ID-E at Advanced Photon Source, Argonne National Lab, USA was utilised here. The experimental setup of micro beam Laue diffraction is represented in Appendix A.

A resolution of ~500 nm was obtained along the incident beam direction (i.e. as a function of depth into the sample) using the Differential Aperture X-ray Microscopy (DAXM) technique. Details about the DAXM technique can be found elsewhere [50] Briefly, the diffraction patterns collected on the detector, is a sum of intensity measured along all points lying along the direction of the incident beam. To resolve the diffraction intensity as a function of depth, a platinum wire was placed between the sample and the detector. This platinum wire moved in sub-microns steps, parallel to the sample surface. The platinum wire act as a differential aperture. By subtracting and measuring the differential intensity corresponding to the sub-micron steps, and extrapolating using the line edge and wire of the beam, a depth resolution of ~500 nm was obtained in the reconstructed data.

<001> oriented grains (~150-200 μm size) were identified across the length of the sample using electron backscattering diffraction (Appendix B). A polychromatic (white) X-ray beam of 7-30 keV was first used to survey the sample and confirm suitable <002> oriented grains were present. The beam was focussed by KB mirrors to a 300 nm × 300 nm spot (FWHM), and measurements were performed in a 45-degree reflection geometry. Detector is placed ~500 mm above sample. Once the target grains were identified, the beam was switched to monochromatic mode for high-resolution depth-resolved strain measurements. In each grain, the strain in the <001> direction was measured using the diffraction setup at 34-ID-E. The measurement gave a strain resolution of the order of $10^{-4}$ [51–53].

For each (00n) reflection, the incident beam was monochromated and the photon energy was scanned to produce a 3D reciprocal space map. The order of the (00n) reflection was chosen to ensure that the diffraction peak centre was in the photon energy range of 11-14 keV. For each reflection, an energy interval of ~30 eV was scanned with 1 eV steps. At each energy DAXM was also carried out to resolve the scattered intensity as a function of depth into the sample. The diffraction data was post-processed using the LaueGo software package [54] and mapped into a 4D space demarcated by the reciprocal space coordinates, $q_x$, $q_y$, $q_z$ and the distance along the incident beam (depth into the sample).

Description of scattering vector is explained elsewhere [55]. **Fig. 1** (b) shows the diffracted intensity, integrated over the tangential reciprocal space directions (normal to the sample surface), and plotted as a function of the scattering vector magnitude (Q) and the depth into the sample.

It can be observed from the surface there is a broad peak in Q (i.e. large variation in strain) for about 2.5 μm, which corresponds with the irradiated layer. Beyond this, the sharp peak drop corresponds to the non-irradiated bulk. From Q, with increasing depth, the transition from the irradiated to the non-irradiated part, a sharp transition in the Q vector magnitude is observed.

This Laue data is used to calculate the lattice strain component in the <00n> direction i.e. normal to the sample surface. The peak centre at each depth (represented by the red dots in **Fig. 1** (b)), ($Q_r$) is found using the centre-of-mass method. Considering small strain approximation, the lattice strain can then calculate as

$$\epsilon_{zz} = \frac{Q_r - Q_{ref}}{Q_{ref}} \qquad\qquad 1$$

Here, $\epsilon_{zz}$ is lattice strain, $Q_r$ and $Q_{ref}$ are scattering vector at irradiated and non-irradiated region of the sample, respectively. $q_{ref}$ is the peak position for the reflection in an unstrained crystal. Here we calculate $q_{ref}$ by considering the average of the peak position in the last 1.5 um depth (which is about 10-12 um away from the surface).

### 2.4. Nanoindentation

Nanoindentation was conducted on Agilent G200 Nano-indenter equipped with a 5 μm spherical diamond indenter tip and a standard Berkovich tip to measure the mechanical properties of ion-irradiated FeCr alloys.

Spherical Indentations were performed to a depth of 200 nm on <001> oriented grains, identified on the samples at different points along the temperature gradient (Appendix B). The depth of 200 nm was chosen such that the plastic zone induced by the deformation is fully contained in the irradiated layer which is predicted to be of depth of about 1.5 μm. A spherical indenter was chosen here to reduce the sensitivity of the measurements to the tip geometry which is particularly evident in cases of sharp Berkovich tips with a high strain gradient near the tip. The experiment utilised the Continuous Stiffness Measurement (CSM) technique, which provided real-time data on stiffness and hardness as a function of depth, supported by a harmonic stiffness parameter with a targeted harmonic displacement and frequency of 2 nm and 45 Hz. A constant strain rate of 0.05 s$^{-1}$ was maintained throughout, while the surface approach velocity was set at 10 nm/s to ensure controlled indentation. These settings minimise the impact of thermal drift and instrument noise, ensuring precise measurements even at shallow depths where surface effects are more pronounced.

Peak loads achieved at a maximum indentation depth of 200 nm across different regions of the sample are compared, and hardness ($H$) is calculated as follows:

$$H = \frac{P}{A} \qquad\qquad 2$$

Where, $P$ is the applied load, and $A$ is the projected area corresponding to the given indentation depth. These measurements do not incorporate corrections for factors such as pile-up, indentation size effects or detailed depth-dependent hardness. A detailed analysis can be made and these are first results.

Berkovich indentation was performed to a maximum depth of 2000 nm on random grains across the temperature gradient using the same nano indenter and CSM parameters (2nm harmonic displacement at 45Hz). The Berkovich tip area function was calibrated against a fused silica standard prior to testing. Hardness and modulus values were averaged over the depth range of 100-200 nm to minimise the roughness effects and to capture mechanical response within the irradiated layer. The strain rate and surface approach velocity were consistent with those used for spherical indentation.

## 2.5. Atomic force microscopy

Atomic Force Microscopy (AFM) was performed using a Dimension D3100 AFM to investigate the surface morphology of indentation. This technique provided high-resolution, three-dimensional topographical data to characterise surface features such as pile-up and sink-in around nanoindentation sites. The AFM imaging complemented nanoindentation results by offering insight into localised deformation behaviour around indents in parts of the samples under different irradiation conditions. Measurements were conducted in contact mode to achieve high-resolution imaging of the indented regions and quantify surface features with nanoscale precision. Scan areas of 10 µm × 10 µm were selected to capture both the indentation and the surrounding surface features. The scanning resolution was set to 512 × 512 pixels, and a scan speed of 1 Hz was maintained to ensure a balance between imaging speed and data quality. AFM provided precise height data, enabling quantification of pile-up heights and distributions around each indentation site. Cross-sectional surface profiles were analysed to identify variations in pile-up and sink-in behaviour across temperature gradient.

All image processing and analysis steps were performed in MATLAB. AFM height data was processed using a custom MATLAB script. Raw surface topography was levelled by fitting and subtracting a plane from the central region of the scan (excluding edge effects), effectively removing sample tilt and background curvature.

Cross sectional profiles were interactively selected along user-defined lines, and the height values were extracted through 2D interpolation along each line. Each profile corresponds to a single line and multiple profiles were aligned based on their minimum height and superimposed for comparative analysis.

All post-irradiation examination, micro-Laue, indentation and AFM were conducted at room temperature.

## 3. Results
## 3.1. Lattice strain

Irradiation-induced lattice strain was measured across a range of temperature points in both Fe5Cr and Fe3Cr using the Laue diffraction setup. **Fig. 2** (a) shows a typical plot of the out-of-plane strain (eq. 1) as a function of depth into the sample. The strain is seen to reduce with increasing depth into the sample, until it reaches a point where it plateaus out in the unstrained crystal below the irradiated layer. From this point the strain is assumed to be zero, this depth is termed as "d0". From our implantation conditions and based on the SRIM calculation (**Fig. 1** (a)), an irradiated layer depth of about 1.5 um was expected. However, in **Fig. 2** (a) it is seen that the strain attributed to irradiation defects extends far beyond that depth (specifically in this case up to 5.8 um) as indicated by the red circle.

For both samples, Fe5Cr and Fe3Cr, a strain curve, similar to the example plot shown in **Fig. 2** (a), was plotted across a range of temperature points. For each case the maximum strain value from this curve is represented as peak strain values in **Fig. 2** (b). The average of the strain values from the surface to "d0" are represented in **Fig. 2** (c). The change in "d0" as a function of changing chromium content and temperature is shown in **Fig. 2** (d). The error bar associated with the measurements in **Fig. 2** (b), (c), and (d) represent the standard error (SE) of the mean, calculated by combining uncertainty from two key sources, uncertainty introduced by the surface detection (available for all temperatures) and reproducibility error from additional measurements taken for selected temperatures at different points on the same <001> grain at a given temperature. The multiple measurements indicate potential sensitivity to local microstructural differences or experimental conditions. A 95% confidence interval is approximated as ±1.96×SE, assuming normally distributed data. SE values and number of measurements for each temperature are provided in a supplementary file. As described in Section 2.3, the strain is calculated from the surface (surface shown in (**Fig. 1** (b)), up to several microns deep into the sample. For some cases, the detection of the surface is more uncertain as compared to others where the surface may be very distinct (Appendix C). For cases where the choice of the surface was unclear, the analysis of the strain measurement results the average of the analysis performed with the multiple choices of the possible surfaces. For some temperature points, multiple diffraction measurements were also analysed. The resulting peak and average strain are represented as error bars in the **Fig. 2** (b) and (c) which represents the uncertainty due to both surface choice and measurement repeatability.

From **Fig. 2** (b), comparing the non-irradiated regions, the hot end (> 500°C) has slightly higher observed strain as compared to the strain in the cold non-irradiated part. This difference is not attributed to temperature effects but is likely due to experimental scatter. The average strain is effectively zero, see **Fig. 2** (c). We note here that there was negligible difference in the strain measurements for the non-irradiated regions from both Fe3Cr and Fe5Cr samples, hence these points are shown collective as single black point. Thus, the non-irradiated data points in **Fig. 2** (b)-(d) can be considered representative of the non-irradiated regions in both Fe5Cr and Fe3Cr samples.

The irradiation induced strain, however, across all temperature points, is significantly higher than the strain in the non-irradiated regions (both cold and hot) for both Fe3Cr and Fe5Cr. For both alloys, it is seen that the peak strain initially decreases as the irradiation temperature rises, reaching almost $0.5\text{-}1.2\times10^{-4}$. Beyond 300°C, the peak strain values start increasing with increasing temperature. The lower chromium content is seen to have slightly higher strain values below 300 °C as compared to Fe5Cr. Beyond 300°C, the peak strain values are seen to be unaffected by the difference in chromium content.

The variation of the average strain with temperature and chromium content (**Fig. 2** (c)) is seen to follow a trend similar to that of the peak strain (**Fig. 2** (b)) and is smaller than peak strain due to averaging across a profile similar to **Fig. 2** (a). For both samples, the average strain values are seen to decrease with increasing temperature till 300°C ($\sim 5\times 10^{-5}$), beyond which the strain values increase with increasing temperature.

Similar to observations in peak strain vs temperature, below 300°C, the average strain values are seen to be higher for 3% Cr than the corresponding values for 5% Cr. Beyond 300°C, the change in the average strain values with temperature is seen to be nearly the same for both chromium content.

**Fig. 2** (d) shows that for both samples, for all the measured points, the depth of zero strain is higher than the depth of the irradiated layer as predicted by SRIM (depicted by the green dotted line in **Fig. 2** (d)). Thus, the lattice strain due to irradiation defects, penetrates well beyond the irradiated layer. The variability in "d0" with temperature is seen to be significantly high below 300°C for both Fe3Cr and Fe5Cr. Beyond 300°C, the "d0" values for the Fe3Cr and Fe5Cr samples, are approximately close to a mean value of 5.25 μm (95% CI, 4.47-6.02 μm) and 4.21 μm (95% CI, 3.93-4.48 μm) respectively. A specific trend in the variation of "d0" with temperature is not seen for either of the two samples. Nevertheless, it is interesting to note that for any given point, the combination of "d0" and the peak strain, captured in the average strain measurement, shows a distinct and similar pattern for both samples.

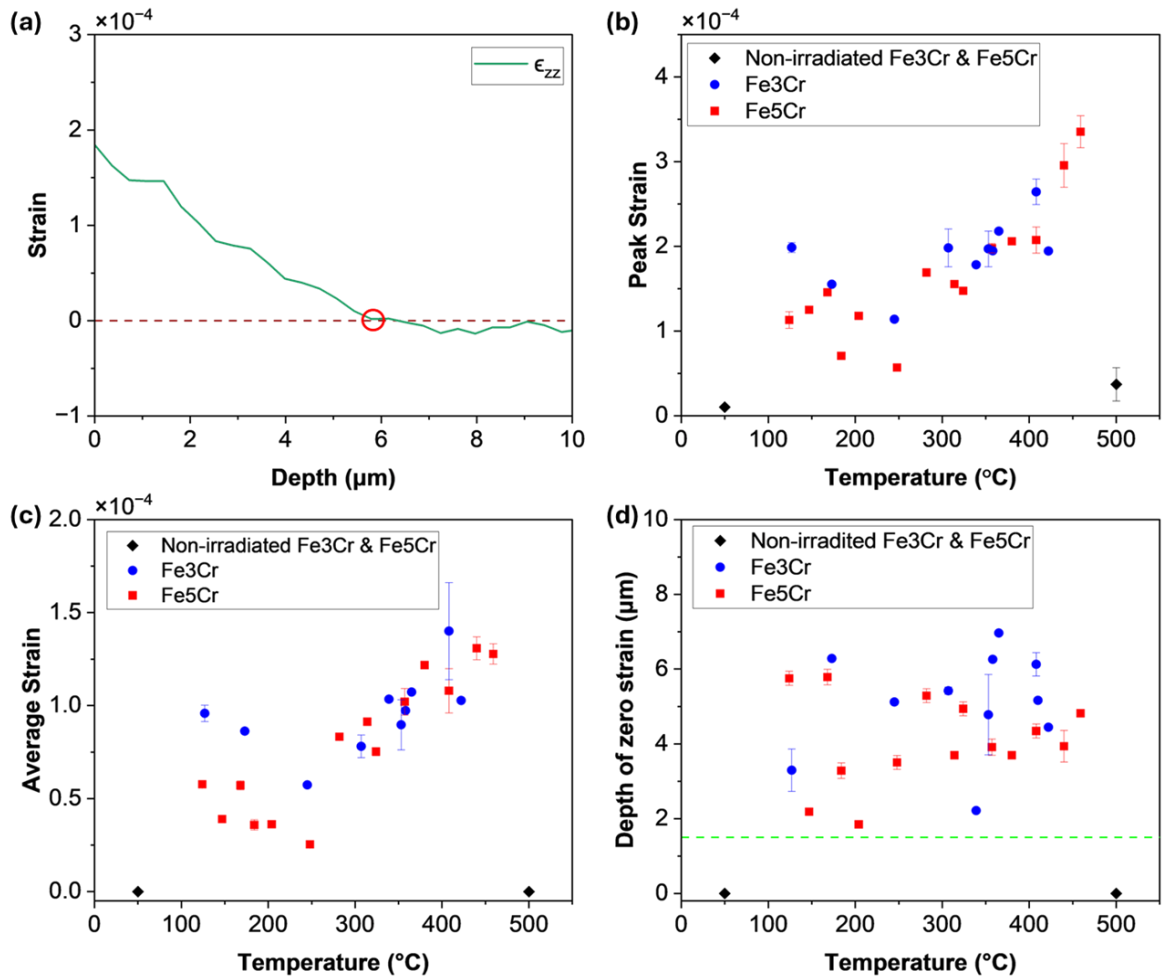

**Fig. 2.** (a) Example of a plot of out-of-plane strain calculated from diffraction measurements plotted as a function of depth into the sample. Here the plotted strain curve corresponds to a measurement at an irradiation temperature of 282°C in the Fe5Cr sample. (b) A plot of the peak strain versus temperature for both Fe5Cr and Fe3Cr. (c) A plot of the average strain over depth "d0" versus temperature for Fe5Cr and Fe3Cr. (d) A plot of "d0" versus temperature for Fe5Cr and Fe3Cr. The dotted green line shows the depth of the irradiated layer as predicted by SRIM. The error bars represent the uncertainty introduced in the measurement due to uncertainty in surface choice (refer to appendix C for details).

### 3.2. Deformation behaviour

We note here that due to lack of suitable grains for indentation, data points could not be obtained for the hot non-irradiated end for the Fe3Cr sample. Berkovich data are represented only for hardness measurements as they are better suited for comparative hardness analysis (**Fig. 3** b-c). Since spherical tips require significantly higher peak loads due to their larger contact area and are more sensitive to the onset of plasticity, the peak load vs displacement is shown exclusively for spherical indentations.

Comparison of the non-irradiated regions in Fe5Cr does not show noticeable change as a function of temperature. However, comparing the irradiated zone to the non-irradiated regions, a non-monotonic trend is observed across the measured temperature range for both Fe5Cr and Fe3Cr. For both, the peak load and relative hardness, initially decrease with increasing temperature up to ~ 300°C, followed by an increase up to ~400°C, and then decline beyond (**Fig. 3** a-c). This behaviour is consistently observed for both spherical and Berkovich indenters, though spherical indenters yield systematically higher hardness values, particularly in the irradiated region.

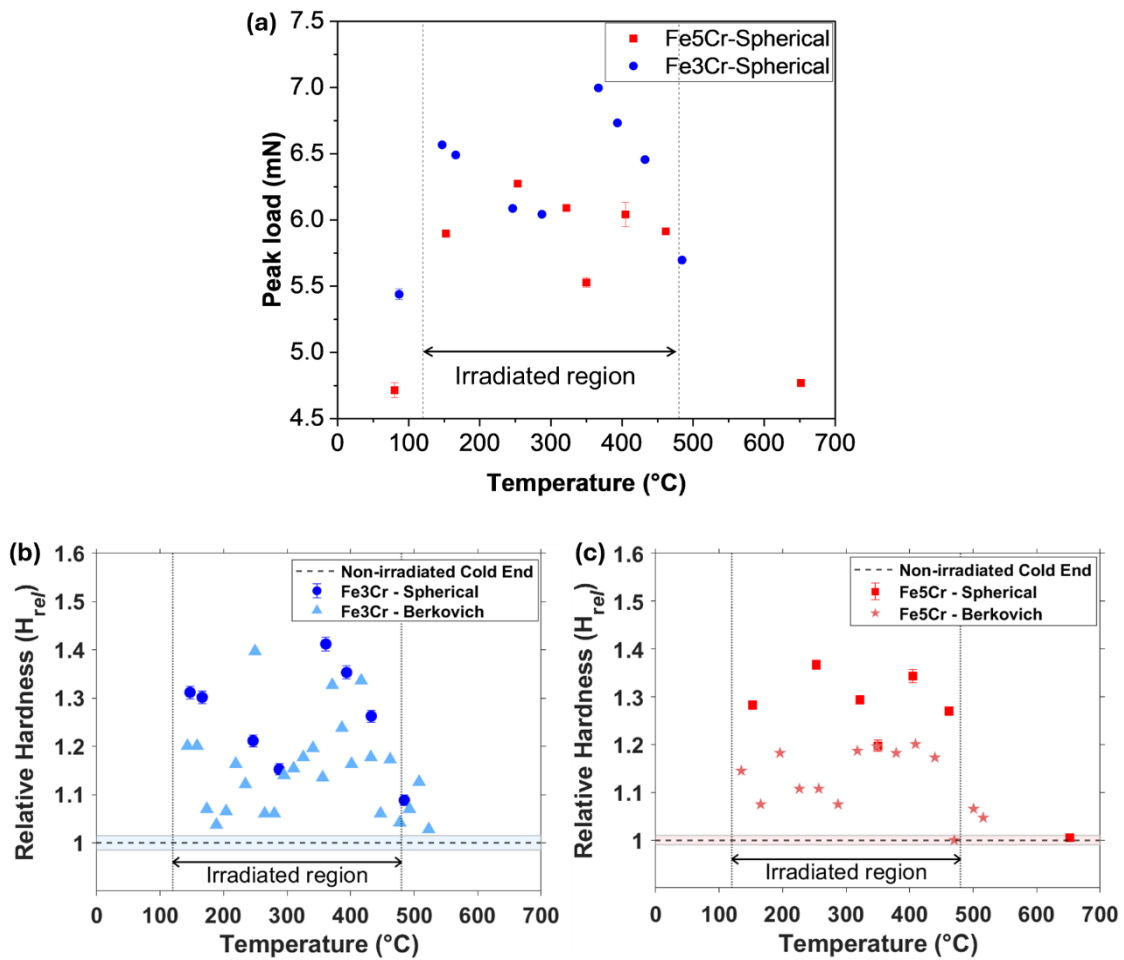

**Fig. 3.** (a) shows the peak load obtained from the load-displacement curves using the spherical tip for both samples as a function of varying temperature points. The error band represents the 95% CI of the repeated indent experiments, detailed analysis is available in appendix D. (b) & (c) shows the hardness (described in eq. 2) as a function of temperature, for Fe3Cr and Fe5Cr, respectively. Error bars represent the uncertainty in the relative hardness, calculated as the ratio of hardness in the irradiated region to that of the non-irradiated reference. For temperatures with multiple irradiated measurements, standard errors of both irradiated and reference hardness were combined using standard error. For temperatures with a single irradiated measurement, error bars reflect only the uncertainty in the reference hardness.

Fe3Cr shows consistently higher peak load and relative hardness than Fe5Cr at corresponding temperatures, regardless of indenter type, indicating greater irradiation-induced hardening in the lower Cr alloy. In case of the lattice strain, the difference as a function of chromium content lowered beyond 300°C which is not seen here for the deformation behaviour. Beyond 400°C, the peak load and hardness are seen to reduce for both Fe3Cr and Fe5Cr, more prominently so for Fe3Cr.

Comparison of the non-irradiated regions shows that the peak load and hardness does not vary with temperature without any irradiation for either of the two samples.

In general, the peak load and relative hardness values for the 3% Cr samples is higher as compared to the 5% Cr sample. For both samples, the peak load and relative hardness varies insignificantly with temperature except in the temperature zone 250-300°C. Outside this temperature zone, the peak load for the 3% Cr and 5% Cr sample is approximately around the average value of 6.47 ± 0.28 mN with a 95% CI and 5.83 ± 0.15 mN with a 95% CI respectively. This is in alignment with the dip in peak strain and average strain values also observed in this similar temperature zone (**Fig. 2**). Across all temperatures, Fe3Cr consistently demonstrates a higher peak load than Fe5Cr, suggesting that the lower chromium content may contribute to stronger resistance to deformation under irradiation conditions.

### 3.3. Surface morphology

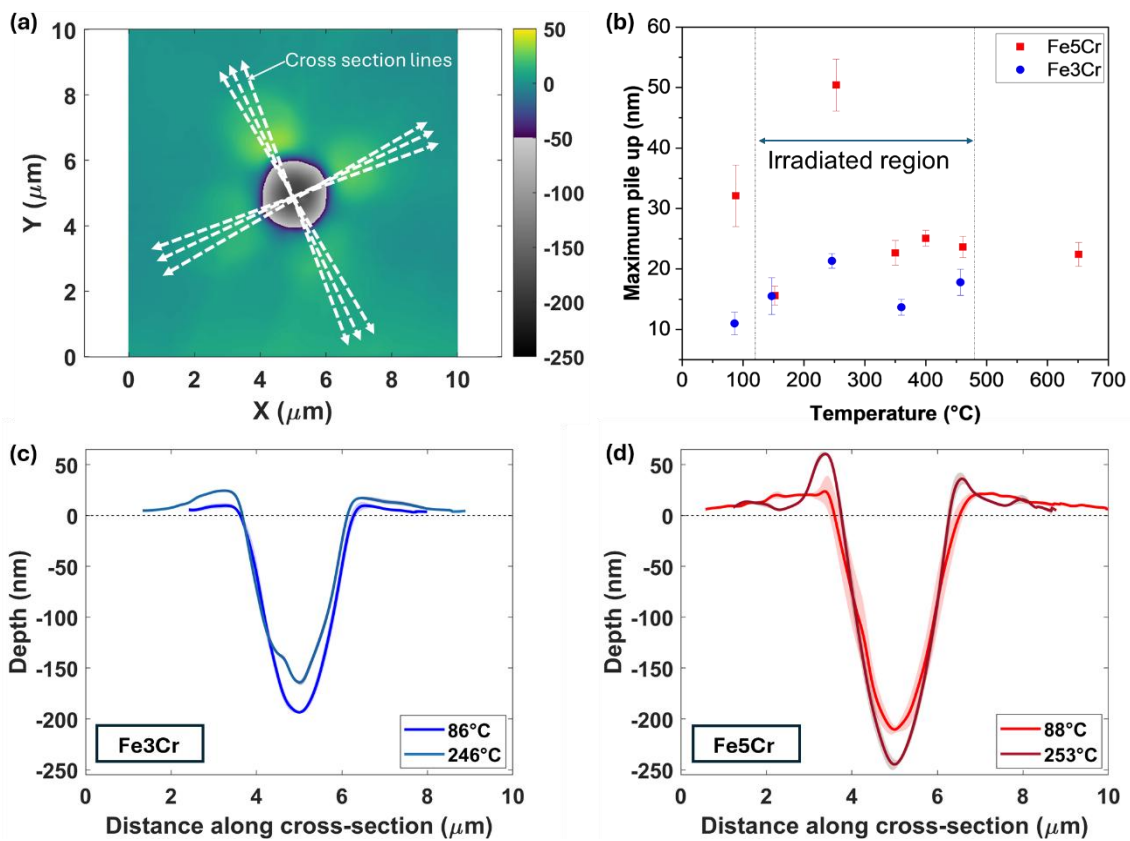

**Fig. 4.** (a) 2D AFM plot of Fe5Cr at 651°C shows six cross-section lines used to extract cross sectional profiles along the pile up. Twelve maximum pile-up heights were obtained from these profiles, and the corresponding standard error was applied in (b). (b) Deformation induced maximum pile up height around indents aligned with the family of {110} plane for various temperatures across the sample, illustrating temperature-dependent variations in material response. (c) and (d) represent the cross-sectional plot of pile up in Fe3Cr and Fe5Cr, respectively, with standard error represented as a shaded band.

A 2D AFM map is shown in **Fig. 4** (a), with white arrows indicating the various cross sections used for analysis. Three cross-sectional lines were taken in each direction to account for possible errors arising from user selection. In total, these six cross-section lines provided twelve measurements of maximum pile-up heights around each indent. An average and standard error were estimated, and maximum pile-up vs temperature was plotted (**Fig. 4** (b)). The cross-sectional profiles of pile up at different temperatures were plotted (**Fig. 4** (c) and d). For **Fig. 4** (c) and (d), the two chosen temperatures correspond to the temperature in the non-irradiated cold end where an AFM measurement was done and the temperature at which the maximum pile-up was observed across each sample respectively (2D plots are available in Appendix E). Demarcation lines are provided in **Fig. 4** (b), representing the temperatures corresponding to the irradiated and unirradiated regions.

Due to the shallow indentation depths, surface pile-up remains minimal across most regions, indicating limited surface relaxation. As shown in **Fig. 4** (b), there is negligible variation in maximum surface pile-up with changes in either chromium content or temperature. The only notable exception is observed in the Fe5Cr sample at 253 °C, where increased pile-up was recorded. However, this region exhibited visible surface oxidation under an optical microscope, which is likely responsible for the anomaly observed in the AFM scan. Additionally, there is little difference in maximum pile-up between irradiated and unirradiated regions, likely due to the very shallow nature of the indents. Overall, under the given irradiation dose and across the studied temperature range, no significant irradiation-induced softening or associated surface localisation is observed.

## 4. Discussion

The results show that even a small irradiation dose of about 0.3 dpa induces a measurable lattice strain in both Fe3Cr and Fe5Cr samples. It is also noteworthy that high-resolution diffraction measurements allow us to probe lattice strains of the order of $10^{-4}$.

Overall, a positive lattice strain is observed for both chromium contents across the entire temperature range investigated. Previous studies indicate that the defect microstructure in self-ion irradiated FeCr alloys is primarily dominated by dislocation loops [56–58]. This aligns well with our results, where the net positive lattice strain can be attributed to the positive relaxation volume associated with interstitial loops [59]. While it remains challenging to deconvolve lattice strain data to directly infer loop size, type and density, our results clearly demonstrate that the defect microstructure evolves notably with changes in both chromium content and temperature.

**Effect of Chromium Content Below 300 °C**

At temperatures below 300 °C, a lower chromium content corresponds to higher lattice strain and higher peak load, indicating a more hardened material. This suggests that even a minor reduction in chromium content can significantly alter the configuration, population, and density of irradiation-induced defects. Our observations align with those reported by Matijasevic et al. [60], where TEM analysis of Fe2.5Cr and Fe5Cr alloys irradiated at 300 °C up to 0.6 dpa (a dose close to ours) revealed larger dislocation loops (~13 nm) well-distributed throughout the matrix in the lower Cr alloy. In contrast, the higher Cr alloy showed smaller defect clusters (~7 nm) with reduced defect density. This disparity in defect microstructure could explain the higher strain and peak load observed in Fe3Cr samples, particularly below 200 °C. Similar findings of smaller loop sizes with increasing chromium content have also been reported [61].

**Effect of Chromium Content Above 300 °C**

Above 300 °C, the difference in lattice strain with respect to chromium content diminishes. This observation is consistent with findings from related studies involving detailed TEM investigations of cavities formed in irradiated FeCr alloys at temperatures between 400–500 °C. These studies [62] show that, while cavities do form under irradiation, the influence of chromium content on cavity formation and associated swelling becomes negligible at higher temperatures. Thus, our observations imply that below 300 °C, chromium content plays a dominant role in shaping the defect microstructure, whereas beyond 300 °C, thermal effects may become the primary driver of defect evolution, thereby reducing the impact of compositional variation.

**Effect of Temperature**

Temperature significantly affects both lattice strain and deformation behaviour, reflecting a substantial change in defect population with temperature. Notably, there is a clear transition in the trend around 300 °C. As temperature increases toward 300 °C, both lattice strain and hardness decrease, and at 300 °C, the impact of irradiation appears minimal—almost comparable to unirradiated material. Beyond 300 °C, however, further microstructural evolution leads to renewed increases in both lattice strain and hardness.

**Low-Temperature Zone (Up to ~200 °C)**

Determining the exact nature of defect configurations in this range requires additional microscopy or atomistic simulations. Nonetheless, it is well-established that room-temperature irradiation of FeCr alloys leads to dislocation loops with Burgers vectors of either $1/2a<111>$ or $a<100>$ [63]. TEM analysis by Prokhodtseva et al. [64] confirms that such conditions produce small loops (~2.5 nm diameter), consistent with early-stage radiation damage. Song et al. [65] report lattice strains of $\sim5\times10^{-4}$ in FeCr alloys irradiated with 0.4 dpa at room temperature. In our study, the observed strains at ~100 °C are about $1-2\times10^{-4}$ (with higher values for Fe3Cr), indicating a reduction in irradiation strain with temperature. This decline in strain continues steadily up to ~300 °C.

**Mid-Temperature Zone (~300 °C)**

Even at 300 °C, the microstructure remains dominated by dislocation loops, with loop size decreasing as chromium content increases [60]. The reduced lattice strain and hardness observed up to 300 °C may thus be attributed to enhanced mobility of these loops. Increased loop mobility can lead to more frequent annihilation at sinks and recombination, resulting in lower defect density and, consequently, lower strain [66]. In situ TEM studies have corroborated this trend of decreasing defect density with increasing temperature [67].

**High-Temperature Zone (>300 °C)**

At temperatures exceeding 400 °C, TEM studies have demonstrated cavity formation in irradiated FeCr alloys, with cavity size increasing with temperature [62]. Although our study does not extend to such high temperatures, this literature indicates a significant microstructural transition beyond 300 °C. This evolution may involve a combination of dislocation loop formation, cavity formation, and Cr precipitation, all of which could contribute to the observed rise in lattice strain and hardening beyond 300 °C. Recent findings by Fedorov et al. [68] show that even trace impurities of carbon and nitrogen can promote Cr clustering at elevated temperatures (>300 °C) through vacancy-impurity interactions. This impurity-assisted clustering may stabilize irradiation-induced defect clusters at high temperatures, providing a plausible explanation for the increasing lattice strain and hardening beyond the transition point. It is worth noting that such contamination (~100 appm of C and N) is common during ion irradiation [69].

**Very High-Temperature Zone (>550 °C)**

Although not directly studied in our experiments, it is reasonable to expect a reduction in irradiation-induced strain at temperatures exceeding 550 °C, in line with earlier reports of irradiation softening in FeCr alloys and steels under high-temperature conditions [70]. For example, the Berkovich indentation at 500 °C show a reduction in irradiation hardening as compared to lower temperatures.

In summary, lattice strain and hardness in irradiated FeCr alloys with <001> orientation show a complex but compelling temperature dependence, particularly in the 200–400 °C range. The observed transitions suggest significant changes in the defect landscape, warranting further electron microscopy to uncover the underlying mechanisms.

**Discussion on AFM Results**

Irradiated metals often exhibit localised plasticity in the form of slip channels and strain localization, especially in grains with <001> orientation [71,72]. This is reflected in the form of surface pile up increase around indents as has been shown for helium-implanted [73]. However, AFM measurements in this study show no significant change in surface morphology or pile-up as a function of either chromium content or temperature. This suggests limited strain localization in these conditions, which can likely be attributed to the low implantation dose and resulting low defect density. This interpretation is consistent with the relatively small lattice strains observed in our samples.

Furthermore, the minimal pile-up detected in AFM scans across all conditions may also result from the geometry of the spherical nano indenter tip. Compared to sharp tips like the Berkovich, spherical tips engage more gradually and generate a wider plastic zone, reducing localized strain concentrations and surface extrusion. This broader distribution of plastic deformation likely suppresses visible pile-up, even in irradiated regions. Hardie et al. [74] observed significant pile-up in Fe12Cr alloys irradiated at similar doses using sharper indenter geometries (Berkovich and cube corner tips) at the indentation depth of 250 nm. In contrast, our study's use of spherical indenter indented up to shallow depths may further be limiting surface strain localisation.

**Discussion on Depth of Irradiation Effects**

An intriguing observation is the depth at which irradiation-induced lattice strain is detected: approximately 4 μm, nearly three times deeper than the 1.3 μm depth predicted by SRIM simulations. Typically, studies conducted at room temperature show good agreement between SRIM predictions and observed strain depth [65]. Our findings suggest that elevated temperatures during irradiation may enhance defect mobility, enabling deeper penetration of irradiation defects. However, this hypothesis requires further investigation using first-principles calculations or atomistic simulations to fully understand the mechanisms driving this behaviour.

Overall, our results clearly demonstrate that even small variations in chromium content, when coupled with temperature, can lead to significant differences in irradiation-induced lattice strain and deformation behaviour in FeCr alloys. These findings underscore the complex interplay between composition and temperature in determining the radiation response of FeCr-based materials, and point to several promising avenues for future research, particularly using microscopy and modelling to further resolve the nature of the evolving defect structures.

## 5. Conclusion

This study investigated the temperature-dependent irradiation response of Fe3Cr and Fe5Cr binary alloys, used as model systems to understand defect evolution in RAFM steels. Employing high-resolution synchrotron Laue diffraction, nanoindentation, and atomic force microscopy, we characterised lattice strain and micromechanical response across a thermal gradient from 120°C to 480°C within each sample.

Lattice strain of the order of $10^{-4}$ was observed at low irradiation doses (~0.3 dpa), with a non-monotonic temperature trend: strain decreased up to ~300°C and increased thereafter. This suggests a transition from mobile, recombining defects at lower temperatures to stabilized defect configurations—such as cavities or solute-decorated clusters, above 300°C. Nanoindentation results mirrored this behaviour, with hardness showing similar temperature dependence. Fe3Cr consistently showed higher strain and hardness than Fe5Cr, especially below 300°C, implying stronger defect retention and possible solute-stabilised clustering.

AFM revealed negligible post-indentation surface morphology changes across all conditions, consistent with shallow indentation depths, reduced strain localization, and diffuse defect structures. Notably, lattice strain extended nearly three times deeper than SRIM predictions, indicating enhanced defect mobility and redistribution at elevated temperatures.

These findings highlight how small variations in Cr content and irradiation temperature strongly influence defect evolution and deformation response. Integrating microscopy (e.g. TEM, APT) with first-principles or atomistic calculations, may help to reveal the underlying mechanisms, enabling a more complete understanding of defect kinetics and interactions in Fe-Cr alloys under irradiation.

**CRediT authorship contribution statement**

**Chandra Bhusan Yadav:** Writing- original draft, Investigation, Conceptualisation, Visualisation, Formal analysis. **Andrew J London:** Funding acquisition, Writing- review & editing, Supervision. **Suchandrima Das:** Funding acquisition, Writing- review & editing,

Supervision. **Tonči Tadić:** Data curation, Methodology. **Stjepko Fazinic:** Data curation, Methodology, Reviewing. **Ruqing Xu:** Data curation, Methodology. **Wenjun Liu:** Data curation, Methodology.

**Declaration of competing interests**

The authors declare that they have no known completing financial interests or personal relationships that could have appeared to influence the work reported in this paper.

**Acknowledgements**

Part of this work has been carried out within the framework of the EUROfusion Consortium, funded by the European Union via the Euratom Research and Training Programme (Grant Agreement No 101052200 - EUROfusion). Views and opinions expressed are however those of the author(s) only and do not necessarily reflect those of the European Union or the European Commission. Neither the European Union nor the European Commission can be held responsible for them. The research used UKAEA's Materials Research Facility, which has been funded by and is a part of the UK National Nuclear User Facility and Henry Royce Institute for Advanced Materials. AJL acknowledges funding from the RCUK Energy Programme Grant No. EP/W006839/1. We thank J. Henry for providing the Fe-Cr alloy material. CBY acknowledges the scholarship support provided by University of Bristol. The authors acknowledge the support of Prof Felix Hofmann and members of his team who supported the experiment at APS.

**Data availability**

Data can be made available upon requesting the authors.

**Appendix A**

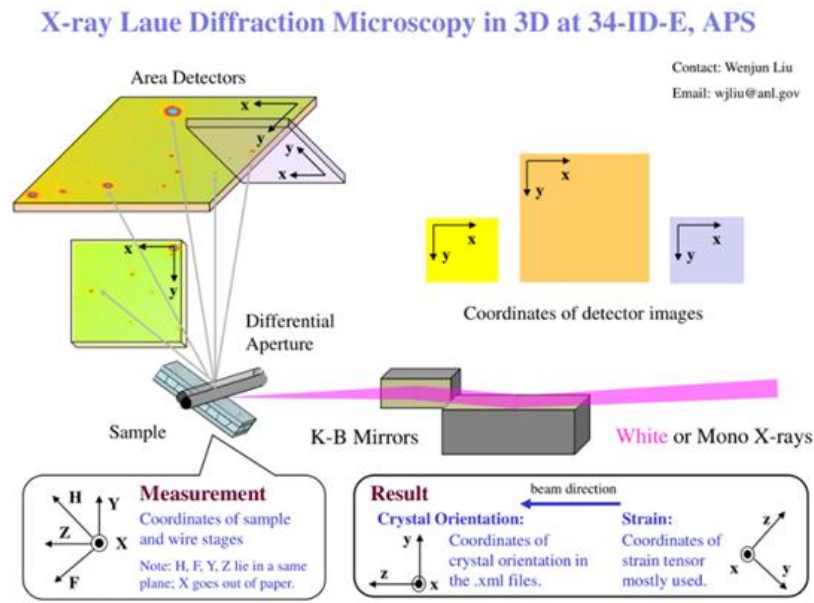

**Fig. A. 1.** Experimental setup pf micro–X-ray Laue diffraction with respective coordinates system [75].

The lattice strain or swelling in the irradiated layer was measured using micro beam Laue diffraction at Argonne National Laboratory, USA on 34-ID-E beamline. The principle behind the Laue diffraction is based on Bragg's law, expressed as

$$n\lambda = 2d \sin \theta \quad \text{Eq. (A1)}$$

Where, $n$ represents an integer, $\lambda$ is the wavelength of the beam, interplanar distance is represented by $d$, and $\theta$ is the angle of the beam. Details of Laue diffraction can be found elsewhere [46]. Briefly, when a beam of an X-ray interacts with the atoms, it diffracts in a specific direction due to diffraction and produces a diffraction pattern. The resulting diffraction pattern i.e. the Laue diffraction pattern consist of peaks that correspond to the constructive interference of waves scattered by the atoms. The different peaks are representative of crystallographic lattice planes. The diffraction pattern being dependent on the interplanar spacing, is altered by change in lattice strain i.e. change in $d$. Thus, by comparing the diffraction patterns of pure and irradiated layer, the distortion of lattice can be measured.

**Appendix B**

The **Fig. B. 1** (a) and (b) represents the matchstick samples of two different chromium content 3% Cr and 5% Cr**,** FeCr alloy. Scorching marks provided during irradiation, are clearly visible on samples. The red grains on the figure below, shows the 001 oriented grains on the Matchstick sample accomplished by electron back scattering diffraction (EBSD).

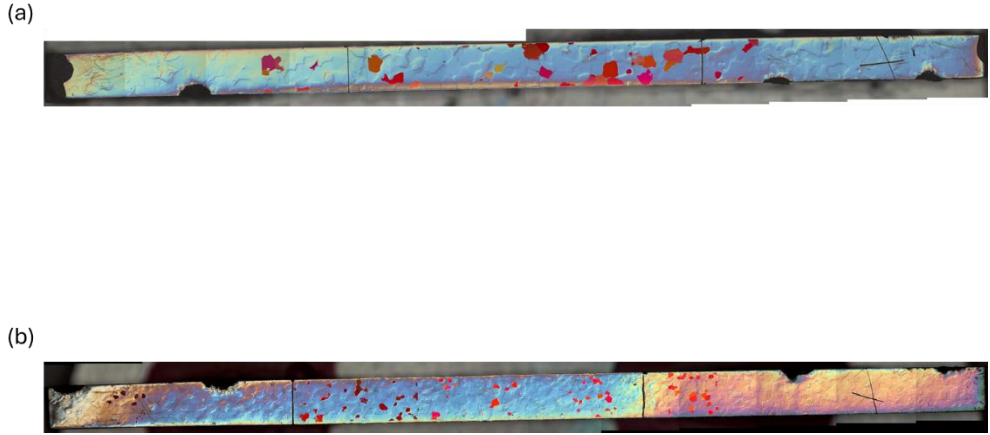

**Fig. B. 1.** Chosen 001 oriented grains from EBSD analysis of (a)Fe3Cr, and (b) Fe5Cr represented in red colour. Figure width is 25 mm.

## Appendix C

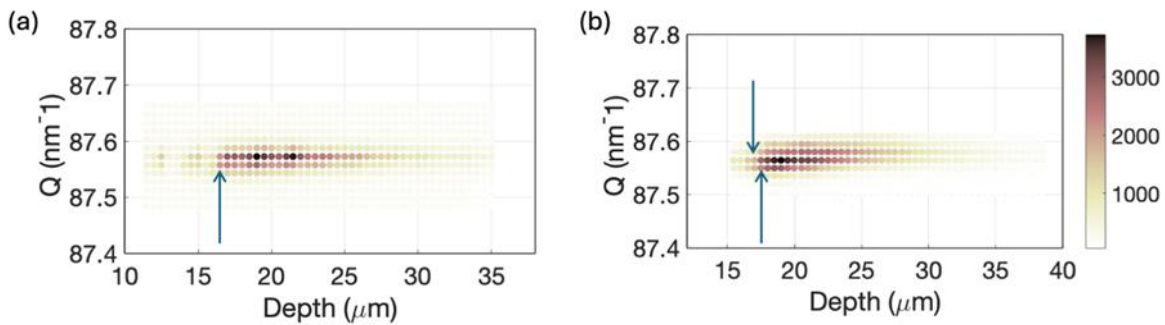

**Fig. C. 1.** Diffracted intensity vs depth showing the uncertainty of surface of the sample.

The above figure shows the diffracted intensity, integrated over the tangential reciprocal space directions, and plotted as a function of the scattering vector magnitude (Q) and the depth into the sample for two cases. In **Fig. C. 1** (a), it is seen that the surface is very distinct as pointed by the blue arrow. However, in **Fig. C. 1** (b) it is seen that the choice of the surface is not very clear. The surface may be at either of the two points indicated by the two blue arrows. So, for the cases, where this kind of uncertainty was present, the analysis of the strain measurement (as described in section 2.3) was done with the multiple choices of surface. These multiple measurements are presented as error bars in the representation of the peak strain or average strain in **Fig. 2** of the main text.

## Appendix D

Reproducibility of nanoindentation experiment was conducted on Fe3Cr and Fe5Cr samples. 001 oriented grains from the cold non-irradiated region of both samples were selected, specifically grain at 80°C and 86°C of Fe5Cr and Fe3Cr, respectively. A 3×3 array of indents was performed on grains with 20 μm spacing between them to avoid interference from the plastic deformation zone. There is an offset between 0.01-0.03°C were applied on temperatures to distinguish among peak loads. A variation of peak loads among repeated array experiments are observed in **Fig. D. 1**. Around 0.5 mN difference is seen between maximum and minimum values for both samples, which comes under the acceptance limit. Mean and standard deviation from these results are calculated and necessary error bar is applied in the **Fig. 3**.

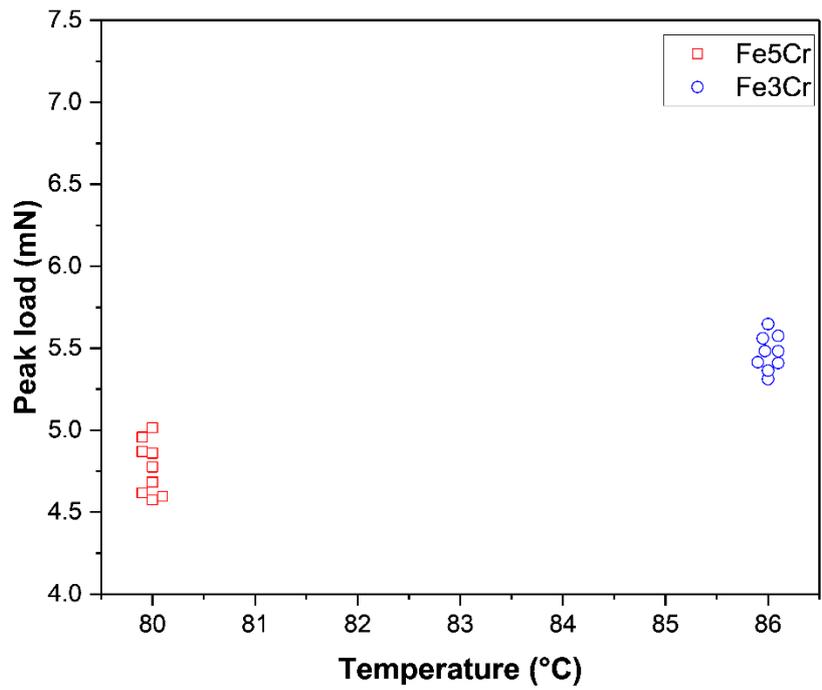

**Fig. D. 1.** Nanoindentation plot of peak load vs temperature of Fe3Cr and Fe5Cr via a 3 by 3 array of indents at 86 °C and 80 °C, respectively.

## Appendix E

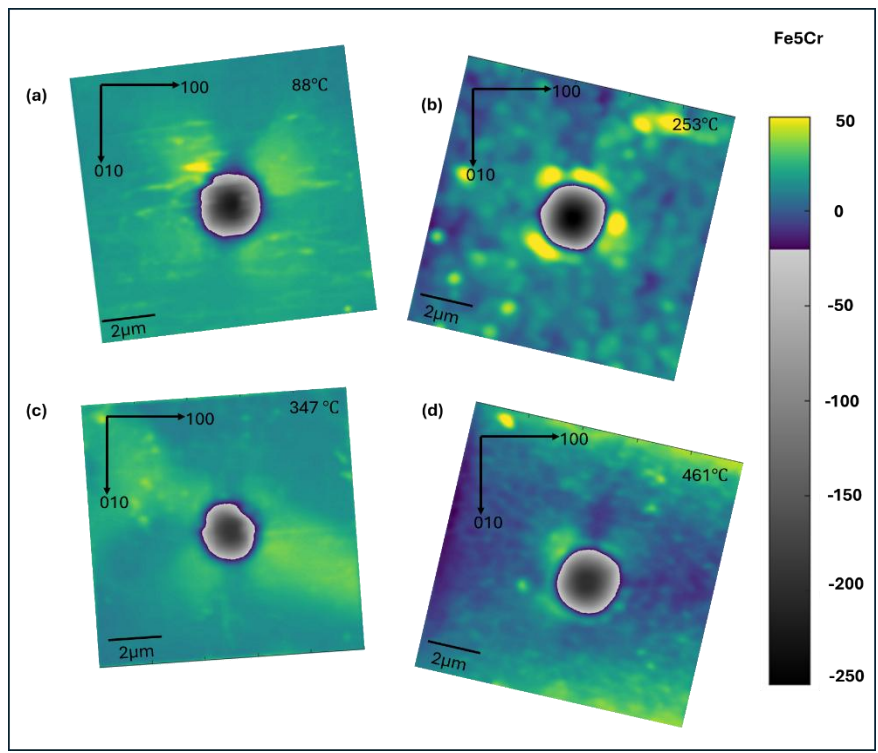

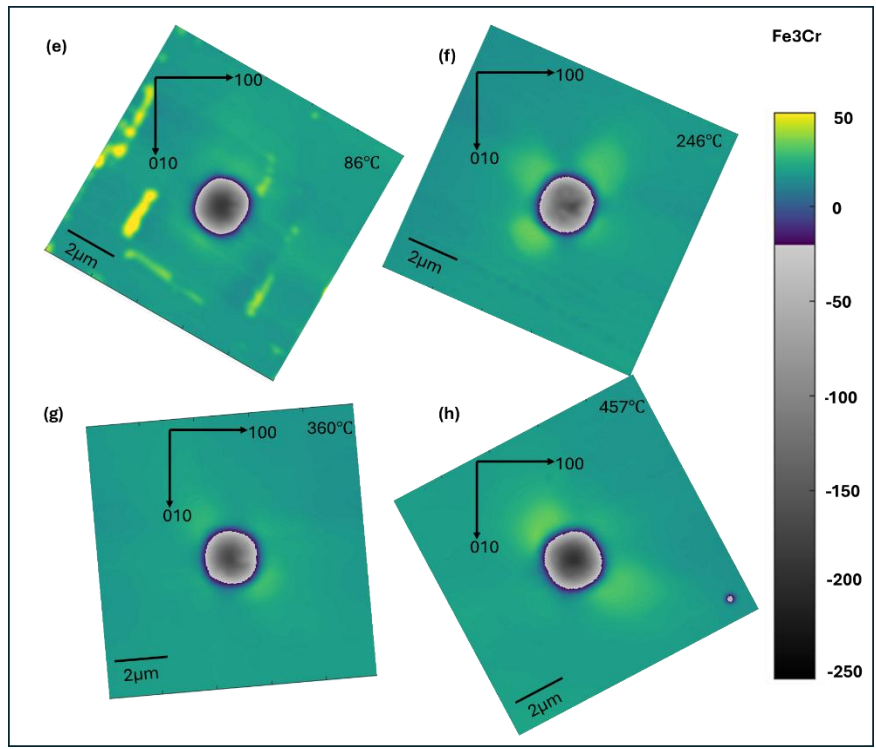

**Fig. E. 1.** 2D AFM plots of indents conducted along the ⟨001⟩-oriented grain at various temperatures. (a–d) and (e–h) represent 2D AFM plots of Fe5Cr (at 88, 253, 347, and 461°C) and Fe3Cr (at 86, 246, 360, and 457°C), respectively.

**Fig. E. 1** Represents the 2D AFM plots of Fe5Cr (a-d) and Fe3Cr (e-h). All images are oriented to align the 100 and 010 directions to horizontally and vertically, respectively. Lighter shade shows the maximum height around the indents, providing visuals of pile-up around the indents. Height is split in gradient of two colours, Viridis colourmap from -50 to 50 nm and grey colourmap from -250 to -50 nm for best visualisation of pile-up. For Fe5Cr, one cold and one hot non-irradiated 2D AFM profile are available, along with two irradiated regions. However, due to difficulties during AFM scanning, only cold non-irradiated indent was scanned for Fe3Cr, which also has three AFM scans of irradiated region.